 \documentstyle[fleqn,12pt]{article}
 \newcommand{\beq}{\begin{equation}}
 \newcommand{\eeq}{\end{equation}}

 \title{Lensing of invisible stars by brown dwarfs}

 \author{Alain Bouquet thanks{Laboratoire de Physique Th\'eorique et Hautes
 Energies, Universit\'es Paris 6 and Paris 7, Unit\'e associ\'ee au
 CNRS (UA 280), 2 Place Jussieu, 75251 Paris Cedex 05 FRANCE}}

\begin{document}
 \vspace{4cm}
 \maketitle
 \vspace{1cm}
 \begin{abstract}

 \vspace{1cm}
 To detect brown dwarfs in the dark galactic halo through
 gravitational lensing, experiments follow the luminosity of
 millions of stars to observe a few lensing events per year. The
 luminosity of a star too faint to be continuously followed can
 be temporarily increased above the detection limit by a
 lensing. The detection of these invisible stars would increase
 the number of events by a factor 2 to 7, and moreover their
 presence would confirm the lensing interpretation of events due
 to continuously monitored stars.
 \end{abstract}

 \vspace{2cm}
 PAR-LPTHE 93-02

 \newpage

 \section{Introduction}

 Two experiments (Bennett et al. 1990, Vidal-Madjar et al. 1991)
 try to detect brown dwarfs in the dark halo of our Galaxy
 (Carr, Bond \& Arnett 1984), through gravitational lensing of
 stars in the Large Magellanic Cloud (LMC). The light of a star
 is amplified when a brown dwarf gets in the line of sight, and
 this amplification varies in a characteristic way with time as
 the brown dwarf moves relative to the line of sight (Paczy\'nski
 1986). Due to the low probability of a lensing event, between
 10$^4$ and 10$^6$ stars (depending on the brown dwarf mass)
 must be monitored for one year to observe one 30 \% luminosity
 increase.

 To follow several millions objects daily, efficient
 algorithms are required to process the huge amount of recorded
 photometric data and to extract the few events looked for.
 First, a pattern recognition algorithm detects the presence of
 some stellar object on a picture, and computes its precise
 position in the sky. Then, its luminosity is extracted from a
 fit of the light distribution, and compared with previous
 measurements. To decrease photometric errors and computer time,
 experiments produce a star catalog from the first pictures, and
 process the following pictures to extract the luminosity curves
 of the stars {\em already} in the reference catalog.

 The drawback of this procedure is that a star absent from this
 catalog (usually a faint star) will never be ``seen''
 afterwards by the detection algorithm, even if it shows up on a
 few pictures, when lensed by a brown dwarf for instance. This
 paper reports the result of a preliminary study, which shows
 that there should be two to seven times more detected lensings
 of invisible stars than of monitored stars.  The number of
 detected lensing events is very sensitive to details of the
 apparatus, to the halo model, to the brown dwarf mass function,
 etc. However, the {\em relative} increase in the number of
 events due to invisible stars depends only weakly on the brown
 dwarf mass, and does not depend on halo parameters or detector
 characteristics. Therefore, it would be a very good test of the
 lensing interpretation of the luminosity variations of some
 stars: these additional events {\em must} be there!

 I therefore suggest to apply the pattern recognition
 algorithm to all pictures, to detect the apparition
 of stars. Because the star catalog was used both to improve
 photometric accuracy and computer time, one could fear that
 my suggestion will be too costly in computer time, and
 inefficient because of the loss in photometric accuracy. This
 does not need to happen, because the only requirement now is
 that a new star appears on a succession of pictures, and then
 disappears. Its precise luminosity does not matter at this
 stage, and therefore the time consuming reconstruction of the
 stellar luminosities is not required. It is only needed
 afterwards, to distinguish a lensed star from a background
 event such as a variable star, but then the star reconstruction
 is only needed for the few candidates, and not for the
 10$^{5\pm1}$ monitored stars.

 \section{Basics of lensing}

 We first recall a few basic notions about micro-lensing
 (Paczy\'nski 1986). When a brown dwarf of mass $M_{{\rm bd}}$
 comes at a distance $R$ to the line of sight of a star, the
 star light is amplified by a factor $A$:
 \begin{eqnarray}
 A &=& \frac{u^2+2}{u\sqrt{u^2+4}} \\
 {\rm where} \;\; u &=& \frac{R}{R_{\rm E}} =
 R \, \left[ \frac{4 G M_{\rm bd}}{c^2} \, \frac{D_{\rm bd}
 (D_{\rm star}-D_{\rm bd})}{D_{\rm star}} \right]^{-1/2}
 \end{eqnarray}
 $R_{\rm E}$ is the Einstein radius, $D_{\rm bd}$ is the
 distance between the observer and the brown dwarf, and
 $D_{\rm star}$ is the distance between the observer and the
 star.

 The event rate $\Gamma(A)$ is the number of times a given star
 is amplified by a factor {\em larger} than $A$ per unit time.
 It depends linearly on the dimensionless impact parameter
 $u(A)$ (Griest 1991):
 \begin{equation}
 \Gamma(A) \,=\, \Gamma_0 \, u(A) \,=\,
 \Gamma_0 \, \left[ \frac{2 \, A}{\sqrt{A^2-1}} - 2 \,
 \right]^{1/2}  \;\;\;\simeq \,\Gamma_0 \, \frac{1}{A} \;\;\;
 {\rm for} \; A \gg 1
 \label{Gamma}
 \end{equation}
 $\Gamma_0$ depends on the brown dwarf mass, on the halo density
 and velocity distributions, and on the direction of the star
 (but does not depend on the nature of the star).

 In the procedure followed by present experiments, an ``event''
 is defined by a minimal amplification $A_0 = 1.34$ of the light
 of a monitored star, the same for all stars. The number $N_{\rm
 monitored}$ of detected events from $N_{\rm stars}$ monitored
 stars during an observation time  $t_{\rm obs}$ then
 writes:
 \begin{equation}
 N_{\rm monitored} = t_{\rm obs}\, \Gamma(A_0) \,
 N_{\rm stars}
 \end{equation}

 For light brown dwarfs, lensing events become too short to be
 detected. The mean duration $t_{{\rm event}}$ of a lensing
 event is:
 \begin{equation}
 t_{\rm event} = \frac{\pi}{2} \, \frac{R}{V_\perp} \;
 \simeq  95 \, 000 \,{\rm s} \;\; u(A_{\rm min})  \, \frac{200
 \, {\rm km/s}}{V_\perp} \left( \frac {M_{\rm bd}}{10^{-4} \,
 M_\odot} \, \frac{D_{\rm bd}}{10 \, {\rm kpc}}
 \right)^{1/2}
 \label{tevent}
 \end{equation}
 where $A_{\rm min}$ is the minimal detectable amplification
 (for present experiments $A_{\rm min}=A_0$). We assume for
 simplicity that all brown dwarfs have the same mass $M_{\rm
 bd}$ and the same transverse velocity $V_\perp$. These
 assumptions can be relaxed, but this would be an unnecessary
 refinement at this level. Possible brown dwarf masses range
 from $10^{-1}\, M_\odot$ (hydrogen burning limit) down to
 $10^{-7}\, M_\odot$ (evaporation limit, De R\'ujula et al.
 1992). The CCD cameras of ongoing brown dwarf searches require
 exposure times of a few minutes to detect stars up to 19th
 magnitude, and the lensing event must appear on several
 consecutive exposures to reconstruct the light curve which is
 the signature of the lensing event. Therefore the minimal
 duration $t_{{\rm min}}$ of a lensing event to be detected is a
 few hours. This leads to a steep drop of the detection
 efficiency for low masses, in which case the only detected
 lensings are due to brown dwarfs slower than average. This
 effect is strengthened by the fact that there is a maximal
 amplification for extended sources, such as red giants, and
 this finite size effect is stronger for lower brown dwarf
 masses.

 \section{Increase in the number of events}

 We are now interested in invisible stars, too faint to be
 detected unless they are lensed. The magnitude $m$ of a star
 amplified by a factor $A$ becomes $m-2.5\log A$. If an
 experiment detects stars up to a limiting apparent magnitude
 $m_{\rm thresh}$, a star of magnitude $m \,\, m_{\rm
 thresh}$ shows up if:
 \begin{equation}
 A(m) = A_0 \times 10^{0.4(m-m_{\rm thresh})}
 \label{Am}
 \end{equation}
 Whereas the minimal amplification is independent of the
 star for monitored stars, it depends on the stellar magnitude
 $m$ for invisible stars. The event rate per star $\Gamma(A(m))$
 then depends on the magnitude $m$ of the lensed star, through
 Equations \ref{Gamma} and \ref{Am}. Large amplifications imply
 short events, therefore there is an upper bound on the
 amplification corresponding to the shortest detectable event,
 and an upper bound $m_{\rm max}$ on the magnitude of a lensed
 star. From Equations \ref{Gamma}, \ref{tevent} and \ref{Am} we
 get for large amplifications:
 \begin{equation}
 m_{\rm max} \simeq m_{\rm thresh} + 2.5 \log \left(
 \frac{95000 {\rm s}}{t_{{\rm min}}} \frac{200 \, {\rm
 km/s}}{V_\perp} \left[ \frac {M_{\rm bd}}{10^{-4} \,
 M_\odot} \, \frac{D_{\rm bd}}{10 \, {\rm kpc}}
 \right]^{1/2}  \right)
 \end{equation}

 The number $N_{\rm invisible}$ of detected micro-lensings of
 invisible stars is a sum over all stars of magnitude
 between $m_{\rm thresh}$ and $m_{\rm max}$:
 \begin{equation}
 N_{\rm invisible} = t_{\rm obs}\,
 \int_{m_{\rm thresh}}^{m_{\rm max}} \Gamma(m) \Phi(m) {\rm
 d}m = t_{\rm obs}\, \Gamma_0  \int_{m_{\rm
 thresh}}^{m_{\rm max}} u(m)\, \Phi(m) \, {\rm d}m
 \label{Ninvisible}
 \end{equation}
 where $\Phi(m)$ is the luminosity function of the target galaxy
 (i.e. $\Phi(m) {\rm d}m$ is the number of stars of apparent
 magnitude between $m$ and $m+{\rm d}m$ in the area surveyed).

 Let us stress that whereas the number of detected lensing events
 (for both monitored stars and invisible stars) sensitively
 depends on detector characteristics, on the normalisation of
 the luminosity function $\Phi(m)$, and, through $\Gamma_0$, on
 the halo parameters and on the mass and transverse velocity
 distributions of the brown dwarfs, all these dependances
 disappear in the ratio $N_{\rm invisible} /N_{\rm
 monitored}$: the number of monitored stars is
 $N_{\rm stars} = \int_{-\infty}^{m_{\rm thresh}} \Phi(m) \,
 {\rm d}m$, and we get:
 \begin{equation}
 \frac{N_{\rm invisible}}{N_{\rm monitored}} =
 \frac{\int_{m_{\rm thresh}}^{m_{\rm max}} u(m)\, \Phi(m) \,
 {\rm d}m}{ u(A_0)\,\int_{-\infty}^{m_{\rm thresh}} \Phi(m) \,
 {\rm d}m}
 \label{ratio1}
 \end{equation}
 If $\Phi(m)$ increases fast enough with $m$, there will be more
 events due to the lensing of faint invisible stars than due to
 bright monitored stars. Moreover, the interpretation of any
 variation of the light of monitored stars as lensing events
 {\em requires} the automatic presence of such lensings of
 invisible stars, which will therefore be a welcome
 confirmation.

 \section{Numbers}

 We take as an example the LMC luminosity function $\Phi(m)$
 given by Ardeberg et al. (Ardeberg et al. 1985) for a small
 area of the LMC, as representative of the mean LMC luminosity
 function. It goes approximately as $\Phi(m) \simeq
 10^{0.4(m-15.2)}$ for $14 < m <23$, which compensates
 the $m$ dependence of $u(m)$ since $u(m) \simeq u(A_0)
 10^{-0.4(m-m_{\rm thresh})}$ for large amplifications.
 Then Equation \ref{ratio1} gives:
 \begin{eqnarray}
 \frac{N_{\rm invisible}}{N_{\rm monitored}} &\simeq&
 m_{\rm max} - m_{\rm thresh} \\
  &\simeq& 2.5 \log \left( \frac{95000 {\rm s}}{t_{\rm min}}
 \frac{200 \, {\rm km/s}}{V_\perp} \left[ \frac
 {M_{\rm bd}}{10^{-4} \, M_\odot} \, \frac{D_{\rm bd}}{10
 \, {\rm kpc}} \right]^{1/2}  \right)
 \label{ratio2}
 \end{eqnarray}
 which depends only logarithmically on the minimal duration
 $t_{\rm min}$ or on the brown dwarf parameters, and gives
 numbers in the range 0-7 for $M_{\rm bd}$ in the range
 $10^{-7}-10^{-1} \, M_\odot$ and $t_{\rm min} = 1 \, {\rm h}$.

 Equation \ref{ratio2} is derived in the limit $A \gg 1$ and
 uses an interpolation of the LMC luminosity function which
 becomes poor for large magnitudes. We can directly compute the
 ratio $N_{\rm invisible}/N_{\rm monitored}$ from Equation
 \ref{ratio1} using actual data and no approximation. Table 1
 shows, as a function of the apparent magnitude $m$, the
 amplification $A(m)$ corresponding to a detection threshold
 $m_{\rm thresh} = 18.9$, the impact parameter $u(A(m))$, the
 factor $u(m) \Phi(m)$ and the duration $t_{\rm event}$. For
 instance, Table 1 shows that lensings of stars of magnitude
 larger than 21 will not be detected if we require a lensing
 event to last more than 3 hours. The durations $t_{\rm
 event}$ given in Table 1 correspond to a 10$^{-4}\,M_\odot$
 brown dwarf with a transverse velocity $V_\perp$ = 200 km/s
 at a distance $D_{\rm bd}$ = 10 kpc, and they scale according
 to Equation \ref{tevent}. The number of monitored stars is the
 sum of the numbers in the second column of Table 1, up to
 magnitude $m=18.9$, that is 38 (in actual brown dwarf searches,
 the surveyed areas are larger and denser than the area surveyed
 by Ardeberg et al., and the corresponding number of stars is
 nearly 10$^5$). The number of invisible stars which can be
 detected when lensed is the sum of the fifth column of Table 1
 between $m=18.9$ and $m=21$ (for $t_{\rm min} = 3$ hours),
 that is 96. The ratio $N_{\rm invisible}/N_{\rm
 monitored}$ in this case is 96/38 = 2.5 .

 This ratio depends on the brown dwarf mass only through the
 $t_{\rm min}$ cut-off, and Table 2 shows how this ratio varies
 as a function of $t_{\rm min}$ for brown dwarf masses $M_{\rm
 bd} = 10^{-1}$ to 10$^{-7}\,M_\odot$. We see that this ratio is
 almost always between 2 and 4, except for very light brown
 dwarfs. The large magnitude bins and the small number of stars
 in each bin induce wild fluctuations in the ratio $N_{\rm
 invisible} /N_{\rm monitored}$ when the threshold $m_{\rm
 thresh}$ or the minimal duration $t_{\rm min}$ are changed
 (this is why we chose a threshold at 18.9 instead of 19).
 Smoothing and interpolation improve the situation, but we saw
 that it does not change the conclusion that invisible stars
 more than double the expected number of lensing events.

 \vspace{0.5cm}
 \noindent {\bf Acknowledgement:} While experimentalists
 involved in ongoing experiments are very busy now, and have no
 time to spare for funny ideas, I wish to thank brown dwarf
 teams for their open mind, and in particular Marc Moniez for a
 first check of the ideas exposed here. So far statistics
 are too limited to give any answer, positive or negative, but
 the test revealed no unexpected difficulty.

 \vspace{0.5cm}
 {\Large {\bf References}}

 Ardeberg A. et al., 1985, A\&A 148,263

 Bennett D.P. et al., 1990, Lawrence Livermore National
 Laboratory preprint

 Carr B.J., Bond J.R. and  Arnett W.D., 1984, ApJ 277,445

 De R\'ujula A., Jetzer Ph. and Mass\'o \'E., 1992, A\&A 254,99

 Griest K., 1991, ApJ 366,412

 Paczy\'nski B., 1986, ApJ 304,1

 Vidal-Madjar A. et al., 1991, 2$^{\rm nd}$ DAEC meeting ``The
 distribution of matter in the universe'' (G.A. Mamon and D.
 Gerbal eds.)

 \vspace{0.5cm}
 {\Large {\bf Table captions}}

 {\bf Table 1:} The first column shows the apparent magnitude
 $m$ of a star in the LMC, the second column is the luminosity
 function given by Ardeberg et al. for a limited area of the
 LMC, the third column is the minimal amplification $A(m)$
 corresponding to a detection threshold $m_{\rm thresh} =
 18.9$, the fourth column is the corresponding impact parameter
 $u(A(m))$, the fifth column the contribution of stars of
 magnitude $m$ to lensing events, and the sixth column is the
 mean duration of the event for a 10$^{-4}\, M_\odot$ brown
 dwarf at $D_{\rm bd}$ = 10 kpc with transverse velocity
 $V_\perp$ = 200 km/s.

 {\bf Table 2:} Ratio $N_{\rm invisible} /N_{\rm
 monitored}$ as a function of the $t_{\rm min}$ cut-off and of
 the brown dwarf mass $M_{\rm bd}$. Beyond $m=23$, the LMC
 luminosity function from Ardeberg et al. was extrapolated by
 $\Phi(m) = 10^{0.36(m-15.2)}$.

 \newpage
 {\Large {\bf Tables}}

 \begin{center}
 \begin{tabular}{|c|c|c|c|c|c|}
 \hline
 Magnitude $m$ & $\Phi(m){\rm d}m$ & $A(m)$ & $u(A(m))$ & $u(m)
 \Phi(m) {\rm d}m$  & $t_{\rm event}$ (hours) \\ \hline \hline
 14.0 & 0 & 1.34 & 1.00 & 0 & 26 \\
 14.5 & 1 & 1.34 & 1.00 & 1 & 26\\
 15.0 & 0 & 1.34 & 1.00 & 0 & 26\\
 15.5 & 1 & 1.34 & 1.00 & 1 & 26\\
 16.0 & 0 & 1.34 & 1.00 & 0 & 26\\
 16.5 & 2 & 1.34 & 1.00 & 2 & 26\\
 17.0 & 7 & 1.34 & 1.00 & 7 & 26\\
 17.5 & 3 & 1.34 & 1.00 & 3 & 26\\
 18.0 & 7 & 1.34 & 1.00 & 7 & 26\\
 18.5 & 17 & 1.34 & 1.00 & 17 & 26\\ \hline
 19.0 & 42 & 1.47 & 0.85 & 36  & 23\\
 19.5 & 69 & 2.33 & 0.46 & 32  & 12\\
 20.0 & 45 & 3.69 & 0.28 & 13  & 7\\
 20.5 & 87 & 5.85 & 0.17 & 15  & 5\\ \hline
 21.0 & 136 & 9.27 & 0.11 & 15  & 3\\
 21.5 & 169 & 14.69 & 0.07 & 12  & 2\\
 22.0 & 300 & 23.29 & 0.04 & 13  & 1\\
 22.5 & 439 & 36.91 & 0.03 & 12  & 1\\
 23.0 & 646 & 58.49 & 0.02 & 11  & 0.5\\ \hline
 \end{tabular}

 {\bf Table 1}

 \begin{tabular}{|r||r|r|r|r|r|r|r|} \hline
 $t_{\rm min}$ & 96 h & 24 h & 12 h & 6 h & 3 h & 1 h & 30 mn
 \\ \hline \hline
 $M_{\rm bd} = 10^{-1} \, M_\odot$ & 2.5 & 3.5 & 4.1 & 4.4 &
 5.0 & 5.4 & 5.9\\
 $M_{\rm bd} = 10^{-2} \, M_\odot$ & 1.8 & 2.9 & 3.2 & 3.9 &
 4.1 & 5.0 & 5.2\\
 $M_{\rm bd} = 10^{-3} \, M_\odot$ & 0 & 1.8 & 2.5 & 2.9 &
 3.5 & 4.1 & 4.7 \\
 $M_{\rm bd} = 10^{-4} \, M_\odot$ & 0 & 0 & 1.8 & 2.1 & 2.5
 & 3.5 & 3.9 \\
 $M_{\rm bd} = 10^{-5} \, M_\odot$ & 0 & 0 & 0 & 0.9 & 1.8 &
 2.5 & 3.2 \\
 $M_{\rm bd} = 10^{-6} \, M_\odot$ & 0 & 0 & 0 & 0 & 0 &
 1.8 & 2.1 \\
 $M_{\rm bd} = 10^{-7} \, M_\odot$ & 0 & 0 & 0 & 0 & 0 &
 0 & 0.9 \\
 \hline
 \end{tabular}

 {\bf Table 2}
 \end{center}

 \end{document}